\documentclass[a4paper,fleqn,usenatbib,useAMS]{mnras}

\usepackage{graphicx}	
\usepackage{amsmath}	
\usepackage{amssymb}	
\usepackage{multicol}        
\usepackage{bm}		
\usepackage{pdflscape}	

\newcommand{\dnu}{\Delta \nu} 
\newcommand{\numax}{\nu_{\rm max}} 
\newcommand{\kepler}{\emph{Kepler} }
\newcommand{\change}[1]{{\color{black}#1}}

\usepackage[T1]{fontenc}
\usepackage{ae,aecompl}

\usepackage{newtxtext,newtxmath}

\title[Asteroseismology of main-sequence F stars]{Asteroseismology of main-sequence F stars with \emph{Kepler}: overcoming short mode lifetimes}

\author[Compton et al.]{Douglas L. Compton$^{1,2}$\thanks{E-mail: d.compton@physics.usyd.edu.au}, Timothy R. Bedding$^{1,2}$, Dennis Stello$^{3}$\\
$^{1}$ Sydney Institute for Astronomy (SIfA), School of Physics, University of Sydney, NSW 2006, Australia \\
$^{2}$ Stellar Astrophysics Centre, Department of Physics and Astronomy, Aarhus University, Ny Munkegade 120, DK-8000 Aarhus C, Denmark \\
$^{3}$ School of Physics, University of New South Wales, NSW 2052, Australia}

\date{Accepted 2019 February 8. Received 2019 February 7; in original form 2018 October 14}

\pubyear{2019}

\begin{document}
\label{firstpage}
\pagerange{\pageref{firstpage}--\pageref{lastpage}}
\maketitle
%
\begin{abstract}
Asteroseismology is a powerful way of determining stellar parameters and properties of stars like the Sun. However, main-sequence F-type stars exhibit short mode lifetimes relative to their oscillation frequency, resulting in overlapping radial and quadrupole modes. The goal of this paper is to use the blended modes for asteroseismology in place of the individual separable modes. We used a peak-bagging method to measure the centroids of radial-quadrupole pairs for 66 stars from the \emph{Kepler} LEGACY sample, as well as $\theta$~Cyg, HD~49933, HD~181420, and Procyon. We used the relative quadrupole-mode visibility to estimate a theoretical centroid frequency from a grid of stellar oscillation models. The observed centroids were matched to the modelled centroids with empirical surface correction to calculate stellar parameters. We find that the stellar parameters returned using this approach agree with the results using individual mode frequencies for stars, where those are available. We conclude that the unresolved centroid frequencies can be used to perform asteroseismology with an accuracy similar to that based on individual mode frequencies.
\end{abstract}


\begin{keywords}
asteroseismology -- stars: oscillations -- stars: fundamental parameters.
\end{keywords}


\section{Introduction}
\label{sec:intro}

The stellar light curves produced from the \kepler mission \citep{koch10a,borucki10} have given an unprecedented insight into the physical properties of solar-like oscillating stars. A number of studies have analysed \kepler light curves and shown that ensemble analysis of main-sequence solar-like oscillators is possible \citep[e.g.][]{appourchaux12,metcalfe14,lund17,silva17}. Future space photometric missions, as well as ground-based spectroscopic observations, will increase the number of possible targets. This papers aims to approach one of the most challenging types of solar-like oscillators, the main-sequence F stars. These stars have large line widths, which makes mode identification difficult \citep[see][and references therein]{white12}.

\begin{figure*}
 \label{fig:figure1} 
 \includegraphics[width=2.0\columnwidth]{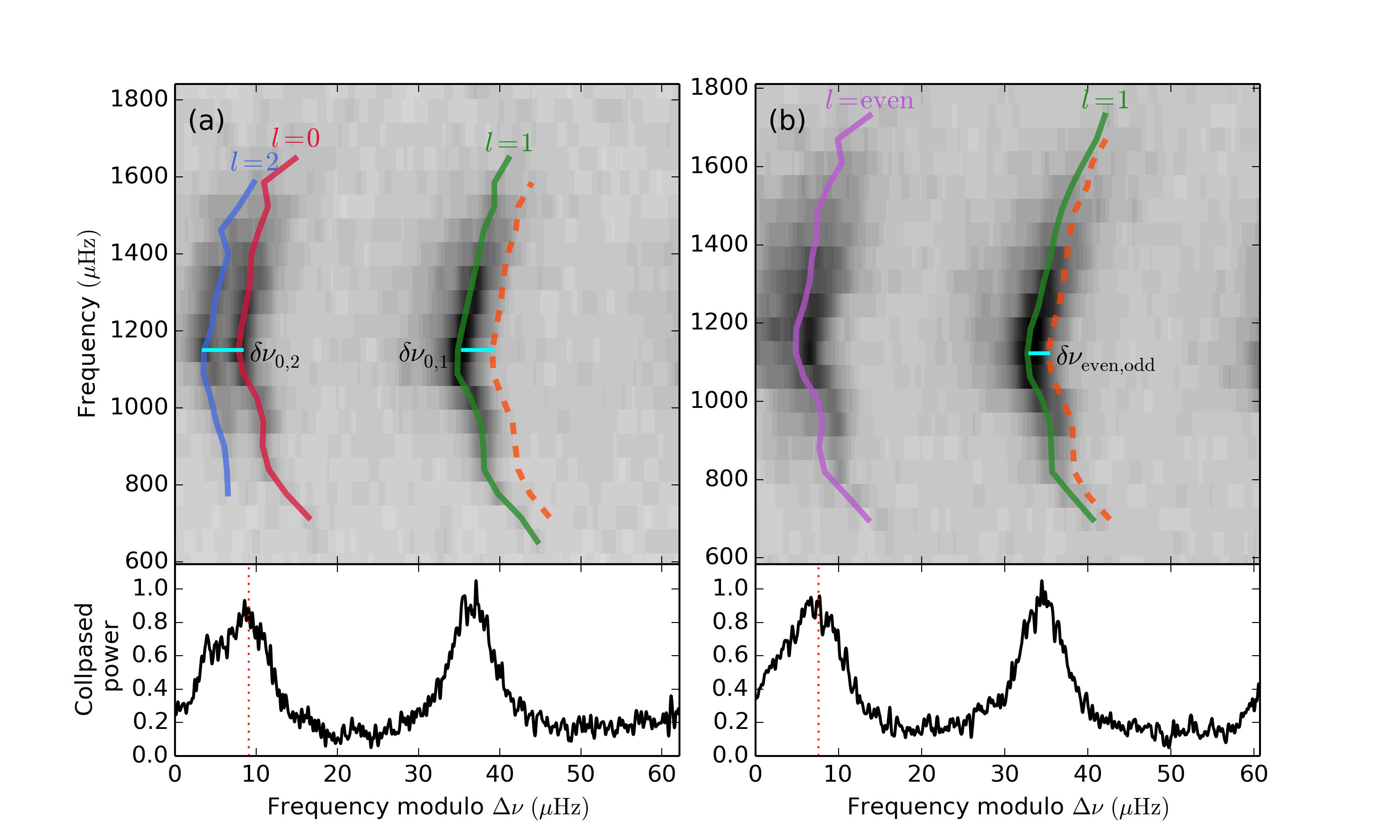}
 \caption{The upper panels shows the power spectra \'{e}chelle diagrams of two \emph{Kepler} stars, classified as `simple' (a, KIC~7510397) and `F-like' (b, KIC~3632418). The greyscale contour represents the Gaussian smoothed (FWHM${=}0.1 \Delta \nu$) power. The ridges of the $l{=}0,1,2$ modes are represented by the red, green, and blue lines, respectively. The purple line indicates the centroid of the unresolved $l{=}0,2$ modes. The dashed orange line is the midpoint in frequency between sequential $l{=}0$ (a) or $l{=}{\rm even}$ (b) radial orders. The lower panels are the collapsed power sum of the upper panels. The red dotted lines represent the value of $\Delta \nu(\epsilon-1)$ for each star.}
\end{figure*}

The oscillation frequencies of solar-like main-sequence stars approximately follow an asymptotic relation \citep[see][]{shibahashi79,tassoul80}:
\begin{equation}
\label{equ:asym}
\nu_{n,l} \simeq \dnu \left( n + l/2 + \epsilon \right) + \delta \nu_{0,l},
\end{equation}
where $\dnu$ is the large separation, $n$ is the radial order, $l$ is the angular degree, $\epsilon$ is a dimensionless offset, and $\delta \nu_{0,l}$ is the small frequency separation between modes of different angular degree with respect to the radial modes. Even in the absence of rotational splitting \citep[e.g.][]{gizon03} radial and quadrupole modes cannot be resolved if the $\delta \nu_{0,2}$ small separation is similar in frequency to the mode line widths. The full-width at half-maximum (FWHM) of the mode is dependent on the damping and is given by 
\begin{equation}
\label{equ:gamma}
\Gamma = \frac{1}{\pi \tau},
\end{equation}
where $\tau$ is the mode lifetime. Solar-like oscillators in which the radial and quadrupole modes cannot easily be resolved are known as `F-like', while unambiguous stars are called `simple' \citep[see][]{appourchaux12}. Fig.~\ref{fig:figure1} shows two examples of \'{e}chelle diagrams.

Due to the unresolved radial and quadrupole modes in F stars, the possibility of accurate asteroseismology is limited. \cite{bedding10} led a multi-site campaign to observe the F4.5 star Procyon and identified oscillation modes in the power spectrum. They suggested that the mode centroids could still be used to do useful asteroseismology. The CoRoT space telescope \citep{auvergn09} also observed a number of F stars including HD~49933 (HR~2530), but unresolved modes hindered analysis \citep[see][]{appourchaux08,benomar09}.

The high-quality \kepler data has since been used to observe individual mode frequencies on some F stars. \cite{guzik16} measured the oscillation frequencies and compare them to stellar models for the brightest F~star in the \kepler field, $\theta$ Cygni, using 2 quarters of data. \cite{lund17} also reported individual frequencies of 22 main-sequence F-like stars included in their so-called LEGACY sample. However, the blended modes make it difficult to be confident that radial and quadrupole modes have been measured realiably in all cases.

We investigate ways to analyse solar-like oscillating F type stars using stellar light curves. The goal of this paper is to confirm the suggestion by \cite{bedding10} that the unresolved $l{=}0,2$ pairs of modes can be useful for asteroseismology. Using the \kepler LEGACY stars, we extract the centroid frequencies from modified power spectra (`simple' stars made `F-like') and calculate stellar and surface correction parameters using the approach of \cite{compton18}. We also revisit a number of non-LEGACY F stars, namely Procyon, HD~49933, HD~181420, and $\theta$ Cyg, using the same methodology to further test the viability of the centroids as seismic probes.

\section{Data}
\label{sec:data}

\begin{figure}
 \includegraphics[width=1\columnwidth]{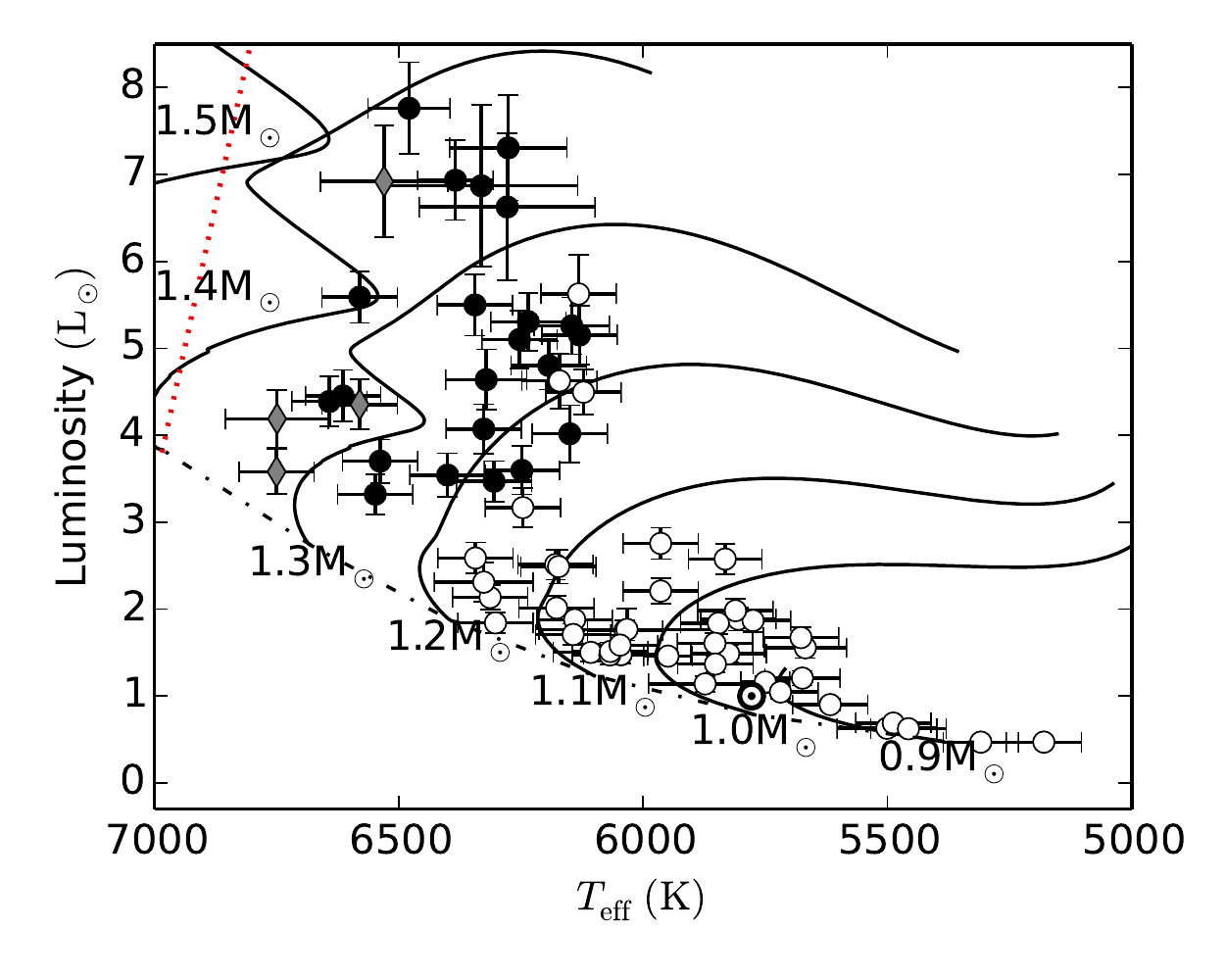}
 \caption{Hertzsprung-Russell (H-R) diagram of the LEGACY stars (circles), four other stars in our sample (grey diamonds), and the Sun ($\odot$). The open and closed circles represent simple and F-like LEGACY stars, respectively. The solid black lines represent the model tracks calculated by \protect\cite{compton18}. The black dash-dotted line indicate the approximate location of the zero-age main sequence. The red dashed line is the cooler boundary to the classical instability strip \protect\citep[see][]{saio98}.}
 \label{fig:figure2} 
\end{figure}

\begin{figure*}
 \includegraphics[width=2\columnwidth]{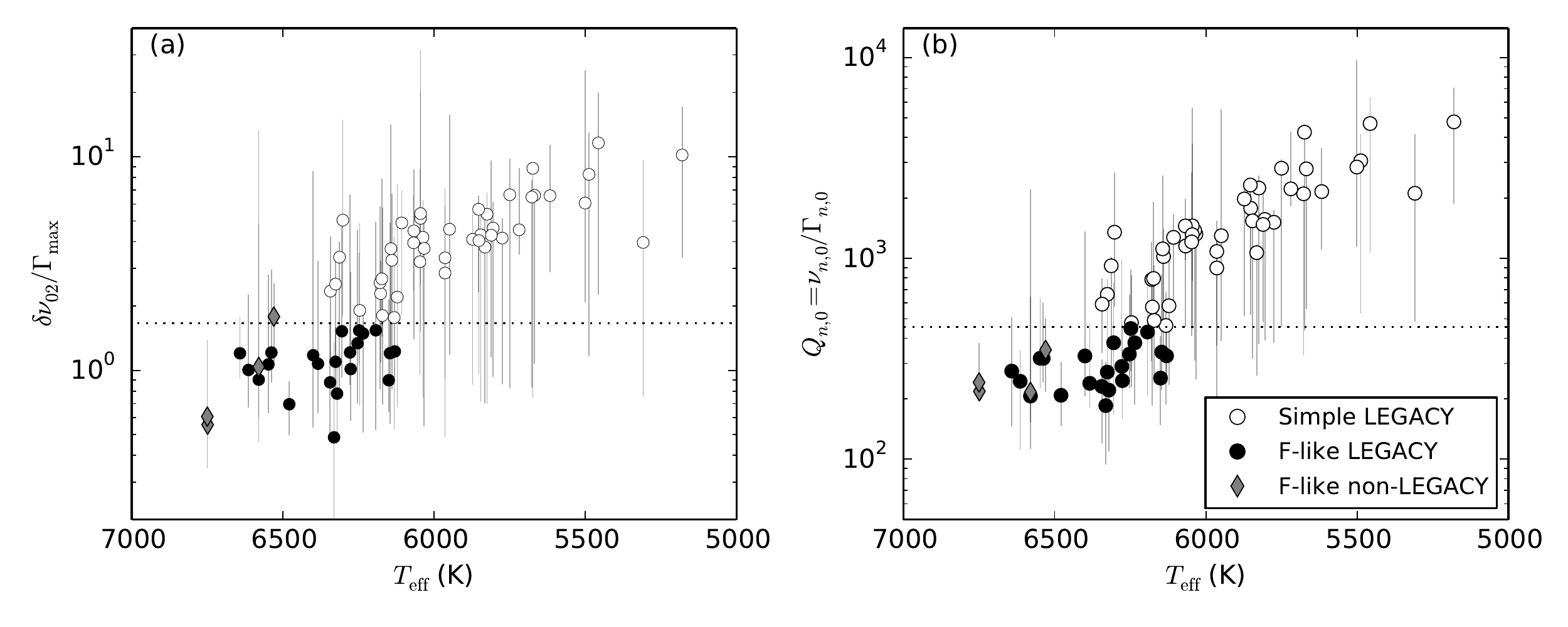}
 \caption{Two measures of how well the modes are resolved, as a function of effective temperature. (a) The ratio between the $\delta \nu_{0,2}$ and $\Gamma$. (b) The quality factor $Q$ against effective temperature for the sample. The symbols are the values at $\numax$ calculated using values of $\Gamma_{\rm max}$ published in their respective papers (see text for details). The symbol shape and colouring are the same as in Fig~\ref{fig:figure2}. The vertical grey lines indicate the range of ordinates using the mode linewidths and frequencies of radial modes for each star. For $\theta$~Cyg and Procyon, individual mode linewidths were not reported and the uncertainty range of $\Gamma_{\rm max}$ was used instead. The black dotted line indicates a boundary between the simple and F-like stars ($\delta \nu_{0,2}/\Gamma_{\rm max} {\simeq} 1.66$, and $Q{\simeq}455$).}
 \label{fig:figure21}
\end{figure*}

The bulk of our sample was taken from the \emph{Kepler} LEGACY sample \citep{lund17,silva17}, which consisted of 22 main-sequence F-like and 44 simple stars that had at least 12 months of short cadence data. Fig.~\ref{fig:figure2} shows their distribution in the H-R diagram.
The boundary between simple and F-like is not quite clearly defined in Fig.~\ref{fig:figure2}. However, there appear to be well defined boundaries, shown in Fig.~\ref{fig:figure21}, when comparing the $\delta \nu_{0,2}$ small separation to the mode line width $\Gamma$ \citep[with the exception of Procyon, see][]{bedding10}, as well as the the quality factor defined by:
\begin{equation}
\label{equ:q}
Q_{\rm max}=\frac{\numax}{\Gamma_{\rm max}},
\end{equation}
where $\numax$ is the frequency of maximum power and $\Gamma_{\max}$ is the mode line width at $\numax$.

The power spectra for the LEGACY sample were obtained from the \emph{Kepler} Asteroseismic Science Operations Center \citep[KASOC;][]{handberg14}. In general, temperatures and metallicities were adopted from the Stellar Parameters Classification (SPC) tool \citep[see][]{buchhave12}. For a small number of stars, temperatures and metallicities were from one of the following: \cite{ramirez09,huber13,casagrande14,chaplin14,pinsonneault12} \citep[see Table~1 from][for details]{lund17}. We considered different effective temperatures of two the LEGACY stars compared to \cite{lund17} \citep[see][for additional details]{compton18}. Our analysis also included four other F stars: the \emph{Kepler} target $\theta$~Cyg, the CoRoT targets HD~49933 and HD~181420, and Procyon. For Procyon we considered both mode identification scenarios. 

For $\theta$~Cyg, \cite{guzik16} extracted the light curve using a custom aperture, and calculated the mode frequencies using Quarters 6 and 8. In our calculation we considered all available \kepler Quarters (Q6, 8, 12 -- 17). We noticed a clear difference in quality between various data releases of the same quarter. Therefore, the data for each quarter was taken from the \kepler data releases which provided the highest oscillation signal-to-noise. Additionally, a manual inspection of each quarter lead to the removal of a number of segments that had greater rms scatter than the rest of light curve. Therefore, the first and last 20 days of Q6, the last 20 days of Q12, and the entire Q16 were excluded in the final light curve of $\theta$~Cyg.

Two F stars observed by CoRoT, HD~49933 and HD~181420, were included in this work. HD~49933 is one of the few stars to have an observed oscillation spectrum from ground-based data \citep{mosser05}. \cite{appourchaux08} calculated the mode frequencies from a 60 day CoRoT light curve of HD~49933, however, their analysis favoured what is now believed to be the incorrect mode classification scenario. This was rectified by \cite{benomar09}, who reanalysed a 180 day light curve of HD~49933 and determined the correct angular degree classification. It has since been one of the most studied F stars in asteroseismology \cite[e.g.][]{kallinger10,salabert11,mazumdar12,liu14}. HD~181420 has also been studied a number of times \citep[see][]{barban09,gaulme09,ozel13,hekker14}. For the analysis of HD~181420, neither scenario was clearly preferred based on the fit of the models to the observed data. However, the correct mode classification has since been clarified \citep[see][]{bedding10b,white12}.

The final star we considered, Procyon, was the first star to have an observed power excess due to the solar-like oscillations, other than the Sun \citep{brown91b}. The oscillations of Procyon have been observed a number of other times \citep[e.g.][]{martic99,bruntt05,huber11b}, but all these studies were unable to resolve individual mode frequencies. We considered data of Procyon from a three week ground-based multi-site campaign \citep[see][]{arentoft08,bedding10}. \cite{bedding10} extracted individual mode frequencies from the observed radial velocity data and noted an ambiguity in the mode identification, with a preference for Scenario B over Scenario A. \cite{white12} used the relationship between $\epsilon$ and $T_{\rm eff}$ to suggest that Scenario B provided the preferred mode classification. However, \cite{guenther14} matched their stellar models more consistently with Scenario A. Note that, Procyon is orbited by a white dwarf in a wide binary, therefore, the mass is well constrained \citep[e.g., see][]{girard00,liebert13,bond15}.

\section{Methodology}
\label{sec:method}

For each star, the background noise was \change{corrected} from the observed power spectrum. The centroid frequencies for the even- and odd-degree ridges were extracted using an MCMC peak-bagging routine. We modified and extended the approach by \cite{compton18} to calculate the asteroseismic and stellar parameters using the centroid frequencies. 

\subsection{Background correction}
\label{sec:bgcorr} 

\begin{figure}
 \includegraphics[width=1.05\columnwidth]{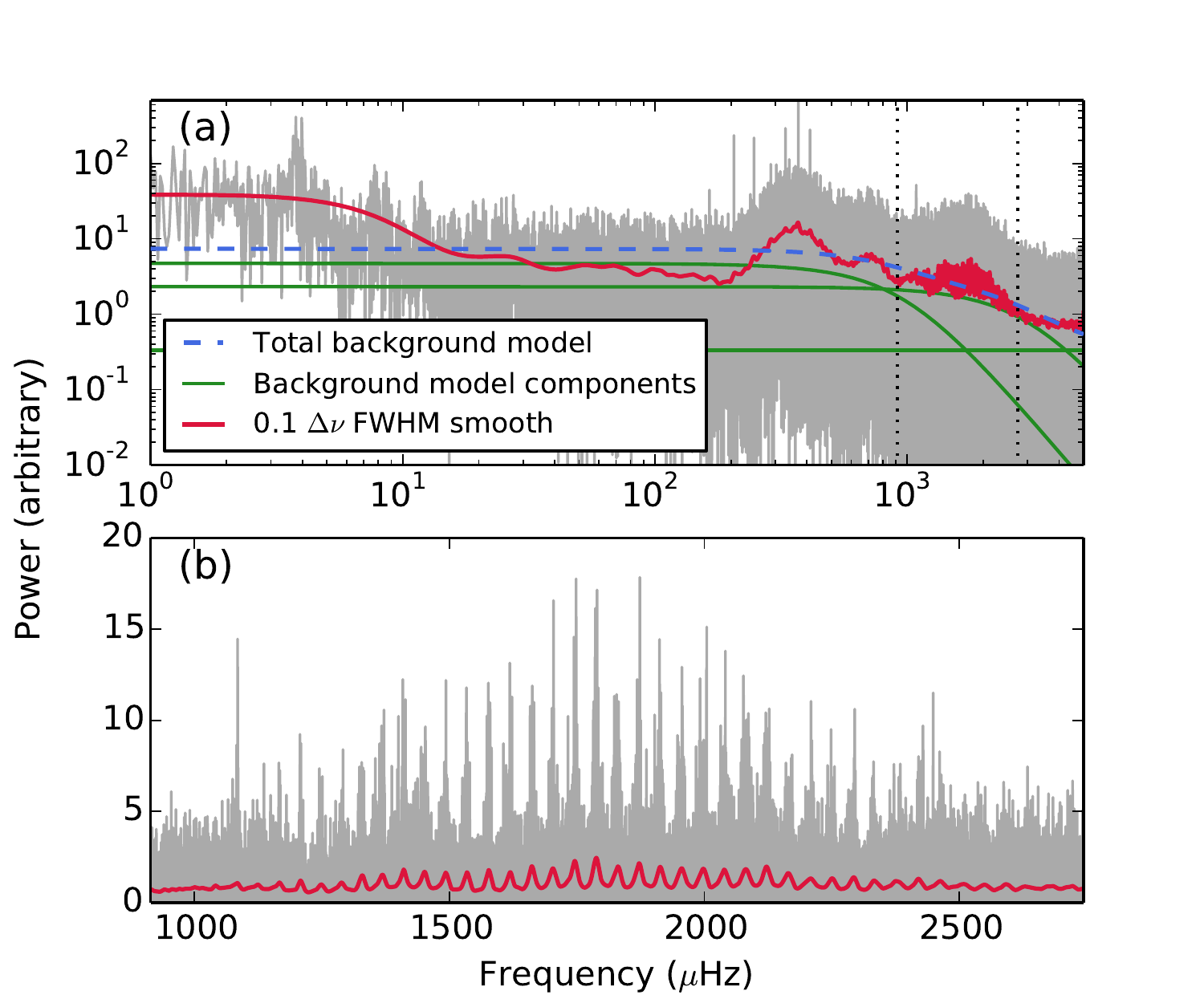}
 \caption{Oscillation power spectrum of $\theta$~Cyg. The grey distribution is the raw power spectrum, and the red line is a Gaussian smoothed version with a FWHM$=0.1\Delta \nu$. The black vertical dotted lines in the top panel (a) indicate the region of the observable oscillations and the frequency range of the lower panel (b). The green solid lines represent the components of the background noise fit defined in Eq.~\ref{equ:harvey}, and the blue dashed line is the sum of the components $\mathcal{H}$. The lower panel (b) shows a close-up of the oscillation modes in the power spectrum after background correction.}
 \label{fig:figure3}
\end{figure}

The noise profile was estimated by \change{minimising $\chi^2$ with 2 degrees of freedom between} a two-termed Harvey function \citep[see][]{harvey93, karoff08} plus a white noise offset and the entire \change{observed} power spectrum, given by:
\begin{equation}
\label{equ:harvey}
\mathcal{H} = \mathcal{W} + 4\sum^{N=2}_i{\frac{\Gamma_i \tau_i}{1 + \left(2 \pi \nu \tau_i \right)^2 + \left(2 \pi \nu \tau_i \right)^4}},
\end{equation}
where $\mathcal{W}$ is a white noise constant parameter, and $\Gamma_i$ and $\tau_i$ are the two components for each term in the function. The power spectrum was then divided by the fitted noise profile to determine the background corrected power spectrum. As an example, Fig.~\ref{fig:figure3} shows the raw and corrected power-spectrum of $\theta$~Cyg, as well as the fitted noise model and its components.

\subsection{Centroid fitting}
\label{sec:rcfitting}

To make a comparison with methods using individual frequencies, we used the same set of modes that were originally extracted from these stars. Therefore, we only considered the radial orders for the even- and odd-degree centroids where the $l{=}0$ and $l{=}1$ modes, respectively, were measured by their respective sources. For example, \cite{guzik16} reported 39 radial and dipole modes in $\theta$~Cyg, therefore we extracted 39 centroid frequencies from the power spectrum. 


To increase our sample of stars with unresolved $l{=}0,2$ pairs, we degraded the power spectra of the simple stars to simulate F-like stars. Each power spectrum was convolved with a Gaussian kernel with a FWHM of $0.1 \Delta \nu$, which approximately corresponds to the minimum width (highest quality factor) of the F-like stars in the LEGACY sample, shown in Fig.~\ref{fig:figure21}. To be consistent, every F-like star was also treated with this smoothing process.

Each mode centroid was measured using a Markov Chain Monte Carlo (MCMC) routine that sampled a one-dimensional Lorentzian function, given by:
\begin{equation}
\label{equ:lorentzian}
L_{n,l} (\nu) = \frac{A\left(\frac{1}{2}\Gamma\right)^2}{(\nu - \nu_{n,l})^2 + \left(\frac{1}{2}\Gamma\right)^2} + 1,
\end{equation}
where $A$ is the height of the mode, $\nu_{n,l}$ is the centroid frequency at radial order $n$ and angular degree $l$ (either even $l=0$ or odd $l=1$), and $\Gamma$ is the line width FWHM. Note that the +1 term in Eq.~\ref{equ:lorentzian} represents the average level of the noise contribution after background correction.

We assumed that the smoothed power spectrum stills follows the statistics of a $\chi^2$ with 2 degrees of freedom. The logarithm of the likelihood function \citep[see][]{duvall86,anderson90,toutain94} is the sum of the log probabilities across all frequency bins $\nu_i$, given by:
\begin{equation}
\label{equ:likelihood}
\ln{\mathcal{L}(n,l)} = - \sum_i{\left[\ln{L_{n,l}(\nu_i)} + \frac{P(\nu_i)}{L_{n,l}(\nu_i)}\right]},
\end{equation} 
where $P$ is the power of the observed spectrum after smoothing and background correction.

Priors were imposed on the free parameters to ensure that they sampled the correct parameter space. Amplitude and line width were trivially bound with uniform priors, $0 < A < \max{(P)}$ and $0 < \Gamma < \Delta \nu / 2$, respectively. A simplified version of the asymptotic relation, Eq.~\ref{equ:asym}, was used to bound the centroid frequency
\begin{equation}
\label{equ:slice}
\left| \frac{\nu_{n,l}}{\Delta \nu} - \left(n + \epsilon + \frac{l}{2} \right) \right| < \frac{1}{2}.
\end{equation}
We neglected rotational splitting of the non-radial modes, since the contribution the non-zero azimuthal components would alter mode profile symmetrically and not affect the centroid frequency. 

The values of $\Delta \nu$ and $\epsilon$ used in our analysis were taken from the same publications as their respective mode frequencies. For completeness, we note that the values for $\Delta \nu$ and $\epsilon$, for use in Eq.~\ref{equ:slice}, can be measured from the power spectrum without the knowledge of individually resolved frequencies. $\Delta \nu$ can be found using an autocorrelation of the power spectrum \citep[e.g.][]{huber09}, and $\epsilon$ can be estimated by solving the asymptotic relation using $\Delta \nu$ and a few individual centroid frequencies. Note that the uncertainty of $\epsilon$ can be large ($\pm 25\%$) because of the wide frequency range of Eq.~\ref{equ:slice}. The only requirement is that the frequency range is narrow enough to include power from either the radial or dipole mode, and not both.

We used the affine invariant MCMC code \texttt{emcee}, written by \cite{foremanmackey13}, to sample Eq.~\ref{equ:likelihood}, which included three free parameters: $\nu$, $A$, and $\Gamma$. We used 200 chains, each sampled 1000 times and initialized at a uniformly distributed value within the parameter space described by the aforementioned priors. Once the sampling was completed, the first 100 samples were discarded to remove some of the burn-in. The median of each sample was used as the final estimation of each parameter. The uncertainty of each parameter was given by the interval spanning 68.27\% of the highest probability density.

\subsection{Model fitting}
\label{sec:mfitting}

We used a similar model fitting technique to that outlined by \cite{compton18}. The same model grid, calculated using MESA \citep[revision 9793,][]{paxton11,paxton13,paxton15} and GYRE \citep{townsend13}, was used as the basis of our stellar models. The inverse-cubic surface correction \citep{ball14} was used to empirically correct the mode frequencies for near surface effects. This included a homology scale factor $r$ to the mode frequencies \citep[see][]{kjeldsen08} to increases the robustness of the fit.

To match the observed data to the models, we adopted the same assumption on the centroid frequency as \cite{bedding10}. That is, we assumed there was a negligible contribution of power from the $l{=}3$ modes, and took the odd-degree centroid as the dipole-mode frequencies. The even-ridge centroid frequencies were modelled as a linear combination of the underlying frequencies:
\begin{equation}
\label{equ:rc_def}
\nu_{n,{\rm even}} = \eta \nu_{n,0} + (1-\eta )\nu_{n-1,2},
\end{equation}
where $\eta$ is the weight of the radial-mode contribution, which is primarily dependent on the power ratios of the corresponding modes. We assumed that the power distribution for each individual mode was symmetrical in frequency. Hence, for a pair of adjacent $l{=}0$ and $2$ modes $\eta$ can be described as a function of mode visibility:
\begin{equation}
\label{equ:W}
\eta \simeq \frac{1}{\tilde{V}^2_2 + 1},
\end{equation}
where $\tilde{V}_2$ is the visibility ratio between the radial and quadrupole modes.

\begin{figure}
 \includegraphics[width=1\columnwidth]{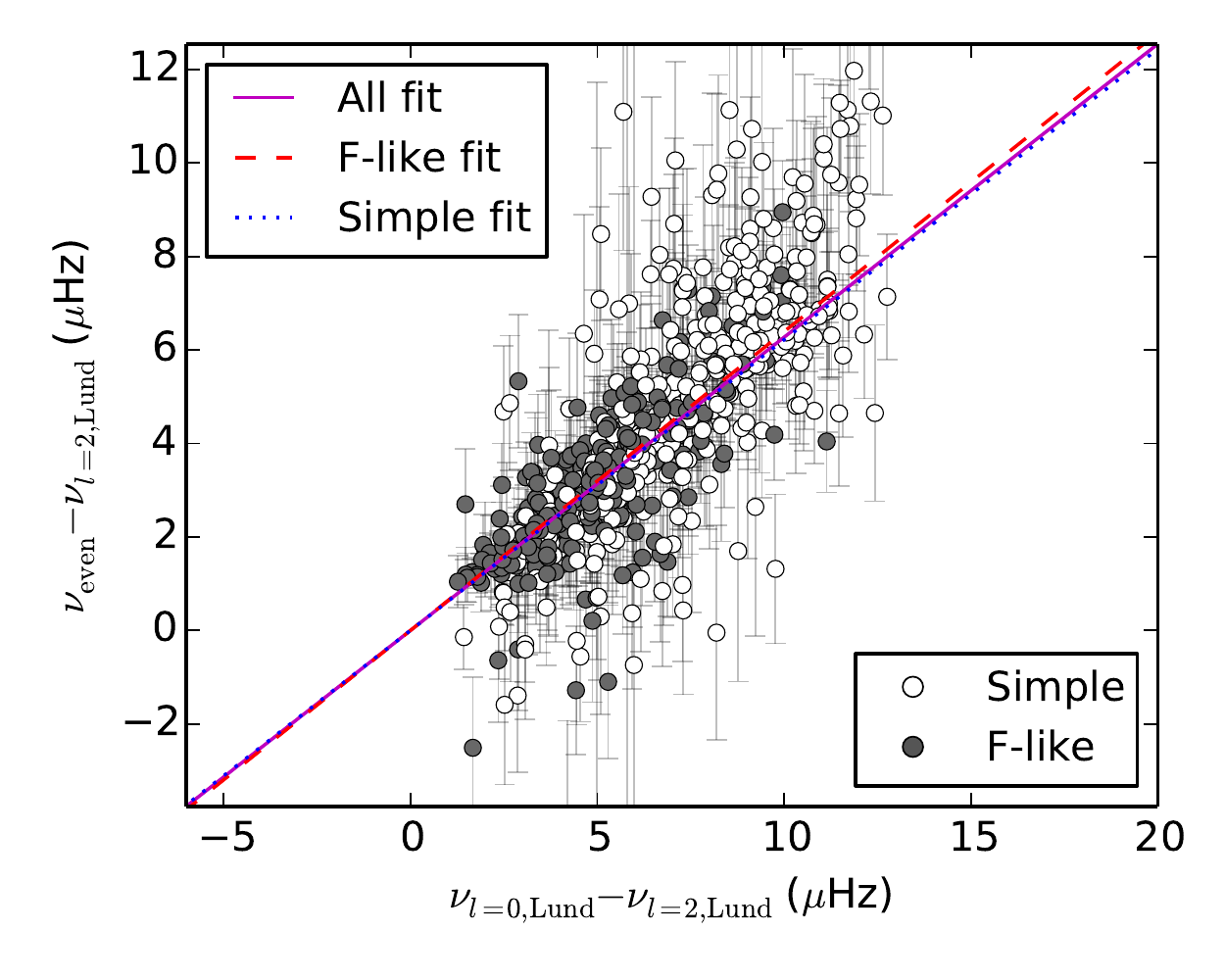}
 \caption{Differences between the measured even-degree centroid frequency and $l{=}2$ frequencies against the $\delta \nu_{0,2}$ small frequency separation, for each mode in the LEGACY sample. The white and grey colouring denotes the simple and F-like stars, respectively. The individual $l{=}0$ and $2$ where taken from \protect\cite{lund17}. The solid purple, dashed red, and dotted blue are the fits to data with Eq.~\ref{equ:eta} using all, the F-like, and the simple stars, respectively.}
 \label{fig:figure14}
\end{figure}

\cite{bedding10} suggested a value of $\eta{=}0.66$ for a star like Procyon if it were observed with {\it Kepler}, based on the results by \cite{kjeldsen08b}. Furthermore, \cite{ballot11} calculated theoretical non-radial visibility ratios for the {\it Kepler} bandpass considering a range of stellar parameters. They found a narrow range of possible visibility ratios for main-sequence solar-like oscillators. To avoid increasing the dimensionality of our grid, we considered a constant visibility ratio for our analysis. We calculated $\eta$ by comparing the individual mode frequencies from each \kepler LEGACY star with the corresponding even-degree unresolved centroid frequencies $\nu_{n,{\rm even}}$ using a linear regression of a rearranged version of Eq.~\ref{equ:rc_def}:
\begin{equation}
    \label{equ:eta}
    \nu_{n,{\rm even}}-\nu_{n-1,2} = \eta (\nu_{n,0}-\nu_{n-1,2}).
\end{equation}
That is, we compared the offset between the even-ridge centroid and published $l{=}2$ frequency for each mode with the $\delta \nu_{0,2}$ small separation. The results, shown in Fig.~\ref{fig:figure14}, show a strong correlation between the two sides of Eq.~\ref{equ:eta}. Most of the outlier even-ridge centroid frequencies originate from the simple stars, seen in Fig.~\ref{fig:figure14}, where there is no ambiguity in mode identification and the frequency can be accurately measured. This is primarily the result of our peak-bagging method failing to accurately measure the centroid frequency, particularly for stars with low signal-to-noise modes.

We found that the contribution of the radial mode to the centroid frequencies $\eta{=}0.628 \pm \change{0.003}$ was less than expected from theory. This value was used to estimate a theoretical centroid frequency when fitting the observed centroid frequencies to the stellar models (see Sec.~\ref{sec:results}). Additionally, the value of $\eta$ did not change significantly when we fit the modes from simple and F-like stars separately (blue and red lines in Fig.~\ref{fig:figure14}). Furthermore, the value of $\eta$ needs only to be as precise as the uncertainty of the mode frequencies. We found in our analysis that using slightly different values of $\eta$ \citep[such as $\eta=0.701$ suggested by][]{bedding10} produced insignificantly different results.

While we used a constant value of $\eta$ throughout this analysis, an exception was made for Procyon, because radial velocity data (the data source for this star) produces different mode visibilities. We adopted the value of $\eta{=}0.49$ suggested by \cite{bedding10} based on the work by \cite{kjeldsen08b} for Procyon.

The remainder of our method follows the \cite{compton18} approach to return the stellar parameters for each of the stars in the sample. This included the application of surface correction when fitting the measured centroid and model frequencies. We assumed that there is a negligible difference between the surface correction of centroid frequencies compared to that of the underlying mode frequencies.

\section{Results}
\label{sec:results}

We split the results into four subsections. Firstly, we continue analysing the results from the centroid peak-bagging. Next we compare the stellar and surface correction parameters returned from the ridge centroid method to the results based on individual frequencies. Thirdly, we present a series of results on the centroid analysis to further show how the ridge centroid can be used. Finally, we discuss the four non-LEGACY stars.

\subsection{Peak-bagging results}
\label{sec:pb_res}

\begin{figure}
 \includegraphics[width=1\columnwidth]{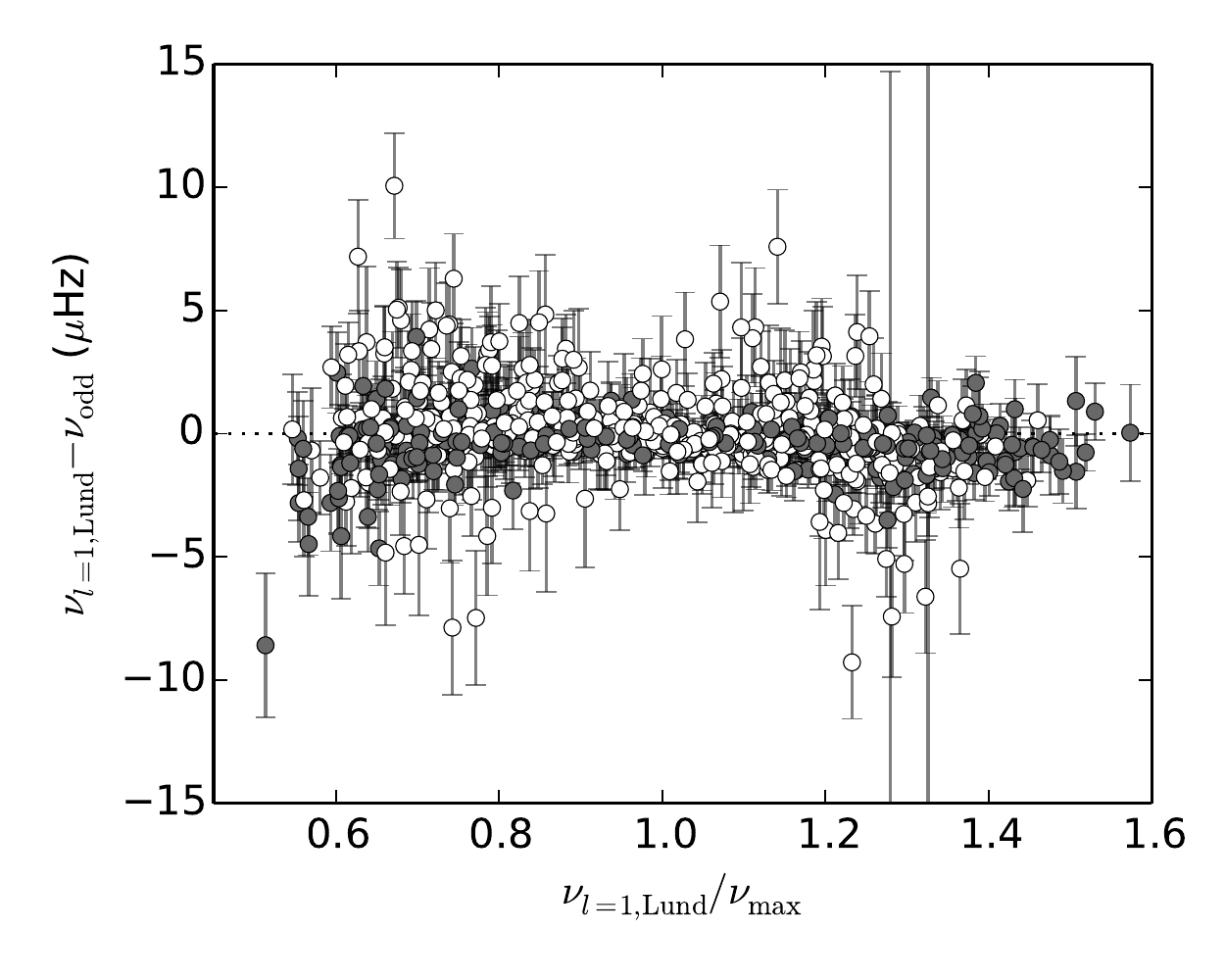}
 \caption{Differences between the measured odd-degree centroid frequencies and the $l{=}1$ mode frequencies published by \protect\cite{lund17}, plotted as a function of dipole-mode frequency normalized by $\nu_{\rm max}$. \change{The white and grey colouring denotes the simple and F-like stars, respectively.} The dotted line marks a difference of zero. }
 \label{fig:figure18}
\end{figure}
To check the validity of our peak-bagging analysis we compared the dipole modes, measured by \cite{lund17}, with the odd-ridge centroid frequencies, which we assumed to be equivalent. Fig.~\ref{fig:figure18} shows the frequency difference as a function of dipole-mode frequency divided by the stars' corresponding value of $\numax$. Modes with a strong disagreement were manually inspected and found to have low signal-to-noise, putting them under a detectable threshold for our peak-bagging method. Modes from the simple stars were more likely to disagree than the F-like stars because the artificial broadening of the modes had a greater impact on their power spectra. The outlier modes, including the ones seen in Fig.~\ref{fig:figure14}, were included in further analysis.

\begin{figure*}
 \includegraphics[width=2\columnwidth]{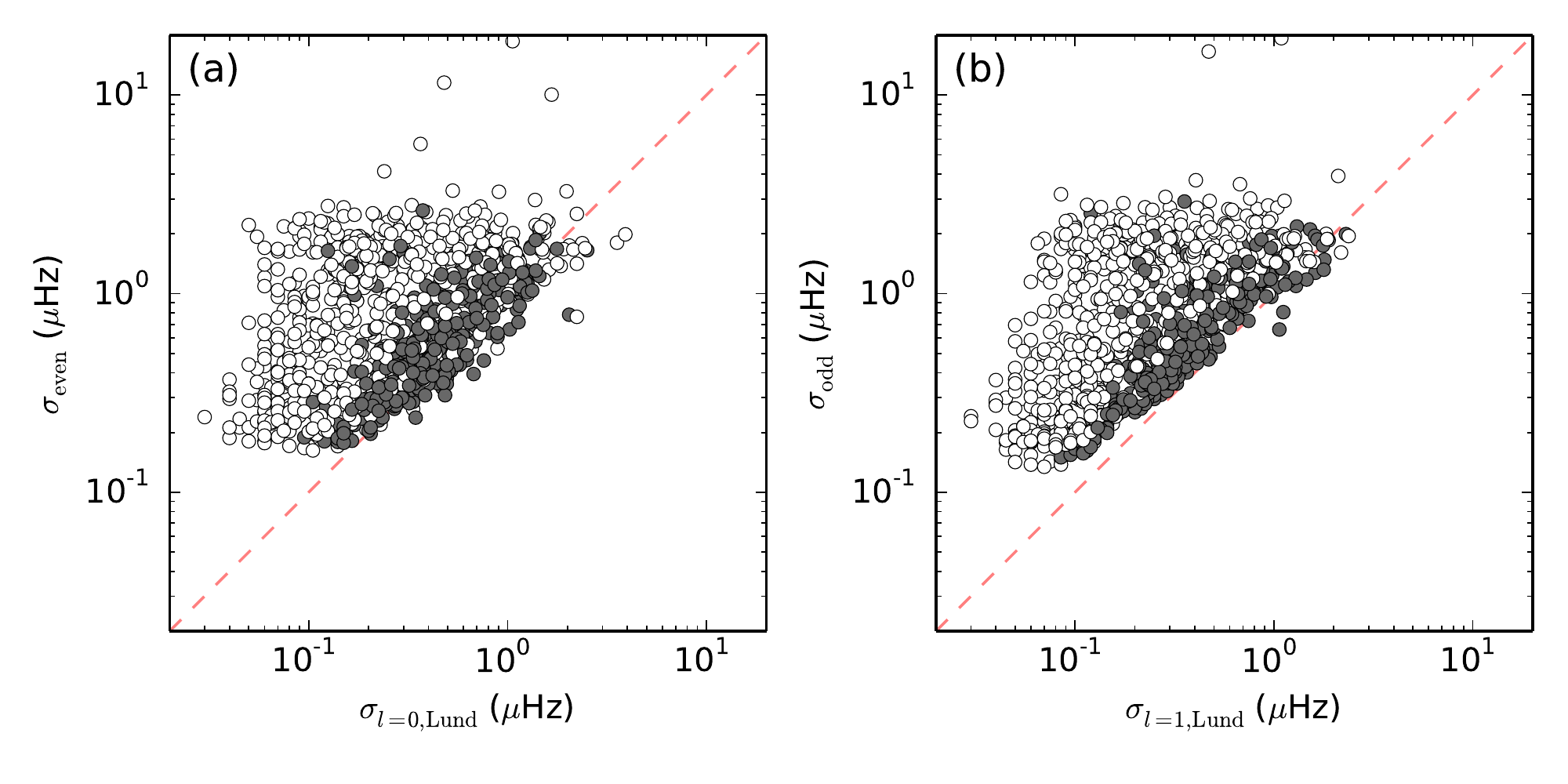}
 \caption{(a) Uncertainties of the unresolved even-degree centroid frequencies compared with the corresponding $l{=}0$ mode frequency uncertainties from \protect\cite{lund17}. (b) odd-degree frequency uncertainties and the $l{=}1$ frequency uncertainties. \change{The white and grey colouring denotes the simple and F-like stars, respectively}. The red dashed lines represent equality.}
 \label{fig:figure19}
\end{figure*}
We also compared the frequency uncertainties with those calculated by \cite{lund17}, shown in Fig.~\ref{fig:figure19}. As expected, the effect of smoothing the power spectra inflates the frequency uncertainties. In general, the uncertainties between the two peak-bagging methods correlate well for the F-like stars. A number of modes have uncertainties many times above the LEGACY value, which suggests the fit did not converge for these modes. The modes with large frequency uncertainties are, once again, dominated by the simple stars.


\subsection{Mode visibility analysis}

\begin{figure}
 \includegraphics[width=1.0\columnwidth]{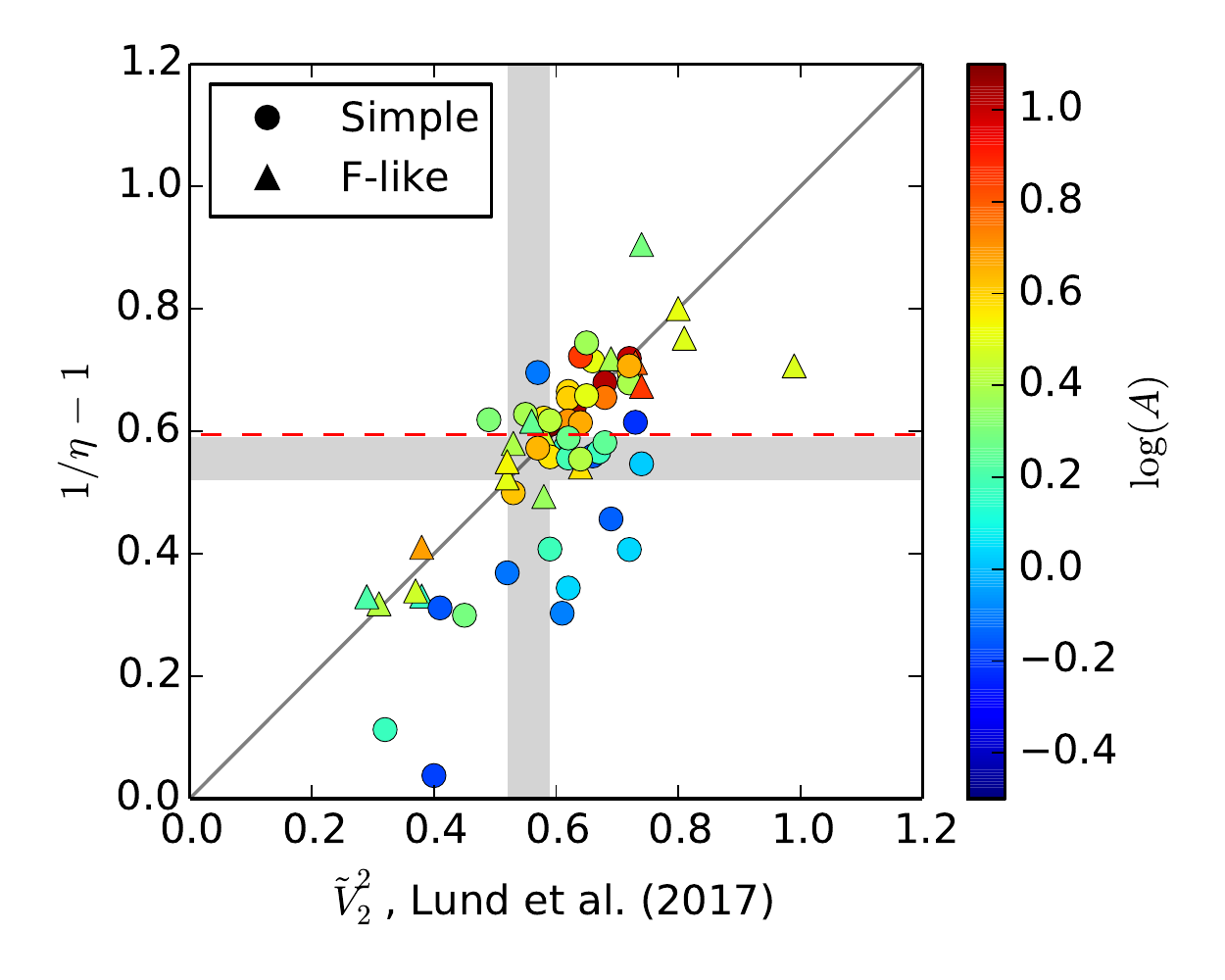}
 \caption{Theoretical quadrupole-mode visibility of the LEGACY stars, defined by Eq.~\ref{equ:W} using the value of $\eta$ calculated for each star using Eq.~\ref{equ:eta}, compared to the quadrupole-mode visibility measured by \protect\cite{lund17}. The circle and triangle symbols indicate if the star is simple or F-like, respectively. The colour of each symbol represents the maximum measured amplitude $A$ of a mode. The shaded region indicates the expected region of $l{=}2$ mode visibilities for main-sequence solar-like oscillators, based on the models calculated by \protect\cite{ballot11}. The red dashed line marks the corresponding chosen value of $\eta=0.628$ calculated from all the LEGACY modes. The solid black line is equality.}
 \label{fig:figure15}
\end{figure}
We estimated quadrupole-mode visibilities for each star using the centroid frequencies relative to the underlying $l{=}0,2$ frequencies. This required fitting Eq.~\ref{equ:eta} to find $\eta$ and converting to a visibility using Eq.~\ref{equ:W} for each star. We compared these values to the visibilities calculated by \cite{lund17}, shown in Fig.~\ref{fig:figure15}. There is a good correlation between the two sets of values, particularly for the modes with higher-amplitude modes. 

Our chosen of $\eta$ was similar to theoretical values calculated using limb-darkening models \citep[see][]{ballot11}, but the visibilities of the individual stars, shown in Fig.~\ref{fig:figure15}, mostly lie outside of the range predicted by \cite{ballot11}. \cite{lund17} gave a number of reasons for this discrepancy which also apply to our work. They also suggested not to assume a fixed visibility in peak-bagging exercises. However, treating $\eta$ as a free parameter would be equivalent to adding a constant term to the surface correction, which would likely cause over-fitting. Creating a function that generates $\eta$ based on fundamental stellar parameters may be possible. However, as mentioned by \cite{lund17}, the mode visibilities correlate poorly with the fundamental stellar parameters. Therefore, we assumed a constant value of $\eta$ based on the fit of Eq.~\ref{equ:eta} to the LEGACY stars, shown in Fig.~\ref{fig:figure14}. In general, this assumption held better for the stars with smaller $\delta \nu_{0,2}$, that is, the F stars.

Note that our value of $\eta$ was based on the LEGACY frequencies, therefore, we expected correlation between $\eta$ and $\tilde{V}$, shown in Fig~\ref{fig:figure15}. The visibilities calculated by \cite{lund17} were based on the relative heights of the modes, whereas ours were calculated under the assumption that the power distribution of the modes was symmetrical. However, the work by \cite{benomar18} may suggest that this assumption is poor, particularly for the higher frequency modes in a given star. An asymmetric mode will shift the ridge centroid relative to the skewness of the asymmetry. \cite{benomar18} did not consider the asymmetry of the modes in F-like stars, where the modes could not be resolved. It should be possible to look for asymmetry in the odd-ridge because the contribution in power should be dominated by the dipole mode.

\subsection{Stellar and surface correction analysis}
\label{ssc_res}

\begin{figure*}
 \includegraphics[width=2\columnwidth]{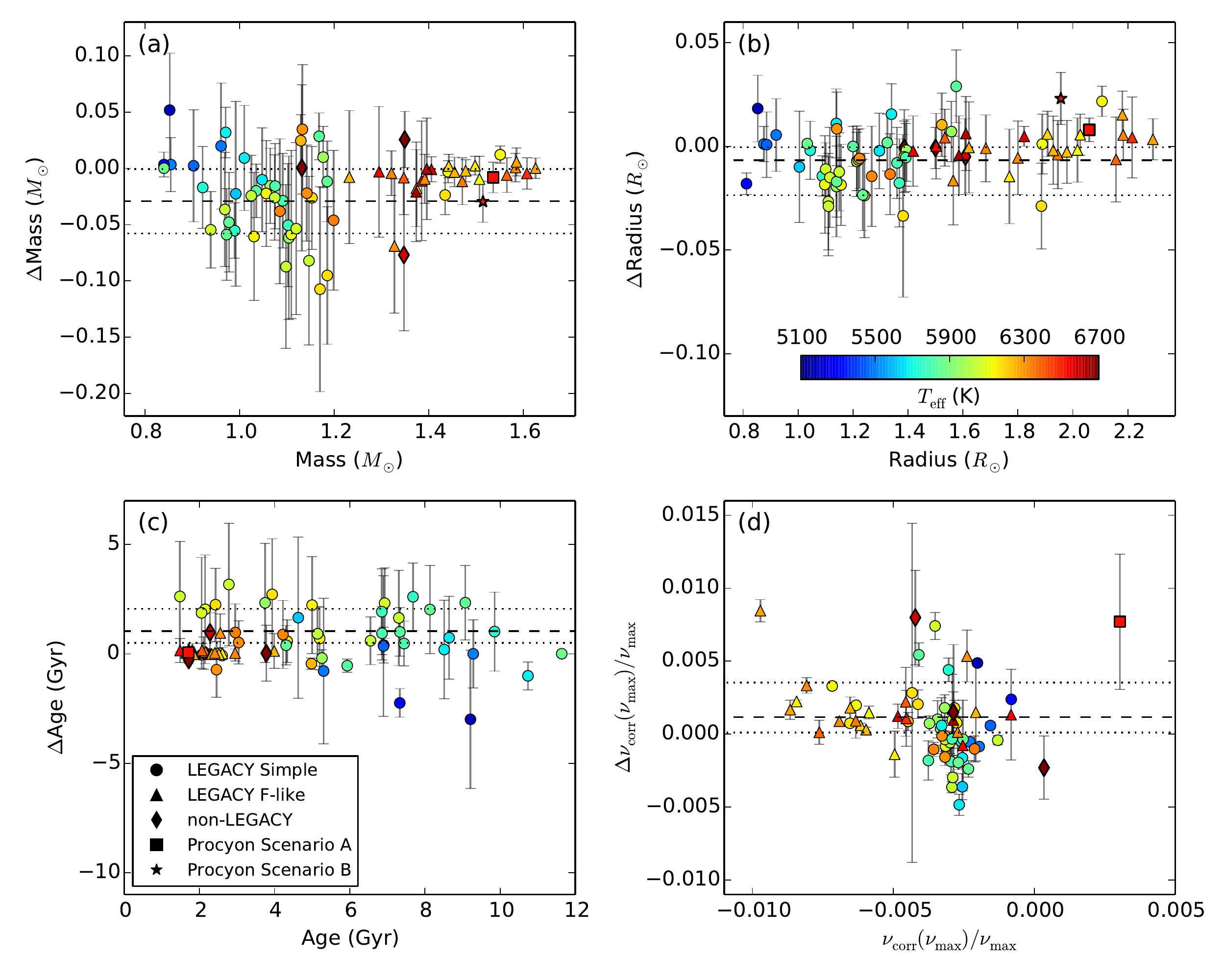}
 \caption{Comparison of the mass (a), radius (b), age (c), and relative surface correction at $\numax$ (d), for the stars in our sample. We plot the absolute differences between the results \protect\citep[using the approach of][]{compton18} from the measured centroid frequencies and using the individual frequencies from their respectively published literature. Differences are plotted as a function of the latter. The symbols denote simple LEGACY stars (circles), F-like LEGACY stars (triangles), the non-LEGACY stars (diamonds), Procyon Scenario A (square), and Procyon Scenario B (star). The colour represents the observed effective temperature of each star, shown in panel (b). The dashed line is the weighted mean of each distribution. The dotted lines is the interval spanning 68.27\% (one sigma) of the data.}
 \label{fig:figure99}
\end{figure*}
Fig.~\ref{fig:figure99} shows the absolute differences for the mass, radius, age, and relative surface correction at $\numax$ between results based on the ridge centroids and those based on the individual frequencies using the \cite{compton18} approach. Our analysis includes both scenarios for Procyon. In general, the methods produced similar results for most stars. 

The variance of the differences is similar to the analysis by \cite{compton18}, where we compared them the \texttt{BASTA} pipeline \citep{silva15} \citep[see Fig.~4 of][]{compton18}. In general, the mass is slightly underestimated for the ridge centroid method, particularly for the simple stars. The calculated stellar ages agree well for the F-like stars between the two methods. In general, the ages of the simple stars are overestimated, up to a factor of $\sim$2, and are less precise overall.

There is a clear difference between the surface correction parameters for the two sets of results. In general, the F-like stars have a smaller correction under the ridge centroid method, while many of the simple stars have a larger magnitude of correction. This makes the relative surface correction at $\numax$, $\nu_{\rm corr}(\nu_{\rm max})/\nu_{\rm max}$, more of a constant across the sample of stars compared the results found by \cite{compton18}.

\if false

\subsubsection{Outlier stars}
\label{sec:outliers}

The outliers in Fig.~\ref{fig:figure99} are predominantly the simple stars. This further suggests that smoothing the power spectra before peak-bagging, to artificially create unresolved $l{=}0,2$ pairs, was not always a suitable practice. After manually inspecting these stars, we found three main reasons which explained most of these discrepancies. 

Firstly, as previously mentioned, our peak-bagging resulted in a small number of outlier mode frequencies where there was low signal-to-noise, particular for the simple stars. These errors were propagated into the calculated stellar parameters. Secondly, the parameter space of the grid of models did not span the entire LEGACY sample, to limit the number of models required for calculation \citep[as dicussed by][]{compton18}. This affected a small number of stars, however, the intention of the homology scaling was to create a model with mode frequencies which could fit these stars. Finally, many of the stars have a theoretical mode visibility ($1/\eta - 1$) that does not agree with our corresponding chosen value of $\eta$ (red dashed line in Fig~\ref{fig:figure15}). Despite these issues, most stars performed well using the unresolved centroid frequency method.

The most obvious simple star outliers are labelled in Fig.~\ref{fig:figure99} and were generally affected by one or more of the issues mentioned previously: KIC~7970740 and KIC~11772920 have a mass outside of the designed grid parameter space; KIC~7106245 has an initial metallicity outside of the designed grid parameter space; the number of radial orders used in the analysis of KIC~8424992 was seven, which is significantly less than the average number; and KIC~9025370 is a known double lined spectroscopic binary, therefore, the metallicity and effective temperature may not be accurate \cite[see][]{compton18}.

\subsubsection{Mixed modes}
\label{sec:mixed_modes}

The centroid frequency method failed to produce an accurate result for KIC~8228742. This is largely due to the implementation of the helium enrichment law in the models. In particular, the strong sinusoidal variations in the mode ridges, seen in more massive main-sequence stars \citep[see][]{verma14b,verma17}, will result in correlated residuals when the helium abundance is not accurate \citep[see][]{compton18}.

Additionally, KIC~8228742 is close to the terminal-age main-sequence, as shown in Fig.~\ref{fig:figure2} ($L{\sim}4.5$ ${\rm L}_\odot$, $T_{\rm eff}{\sim}6100$ K). Stars at this stage of evolution can possess mixed modes, which are coupled p and g modes and exhibit characteristics of both \citep[see][]{osaki75,aizenman77,dziembowski91,cd04,aerts10}. Mixed modes strongly deviate from the asymptotic relation and their frequencies evolve rapidly compared to pure p modes. Therefore, fitting mixed modes to a grid of models is inherently difficult. The stars in the LEGACY sample were specifically chosen to not contain mixed modes, hence, stellar models with mixed modes should be omitted for samples like this.

The value of the centroid frequency, defined by Eq.~\ref{equ:rc_def}, includes information about the modelled quadrupole modes. Therefore, the fit should be somewhat resilient to models with mixed modes due to their significant frequency difference. This does not appear to be the case for KIC~8228742, and the poorly modelled helium abundance is partially at fault. In Sec.~\ref{sec:procyon}, we show that Procyon has a similar issue with quadrupole mixed modes.
\fi

\subsection{Small separation analysis}
\label{sec:rc_res}

\begin{figure*}
 \includegraphics[width=2.14\columnwidth]{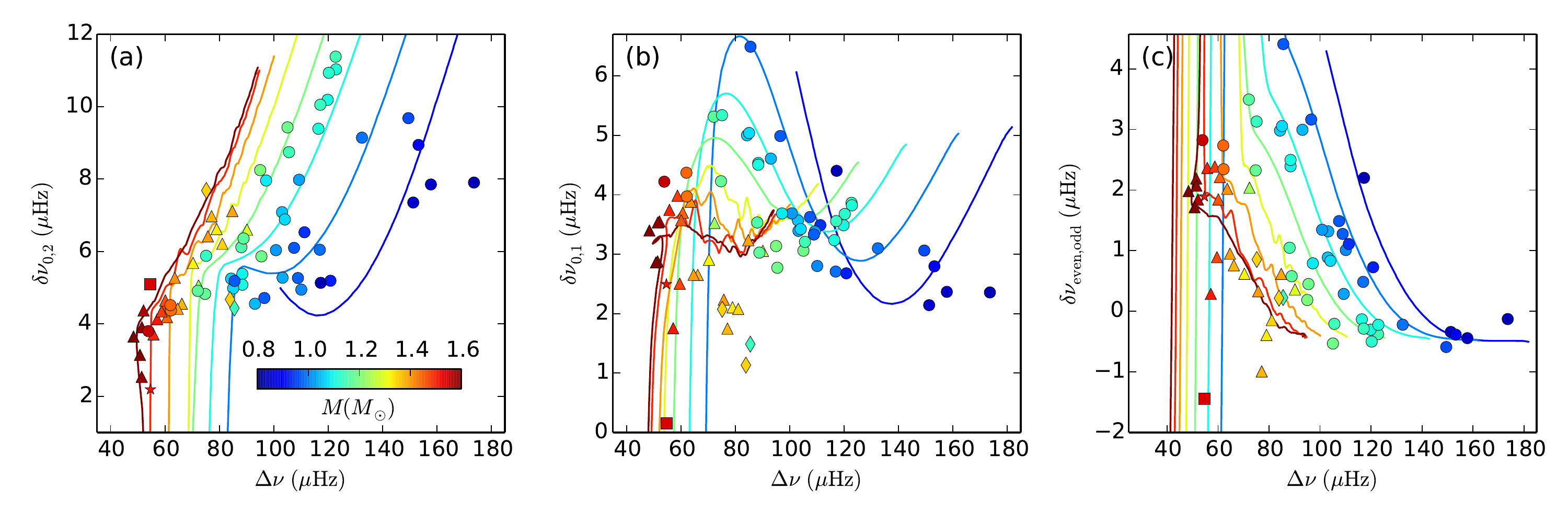}
 \caption{C-D diagrams for the $\delta \nu_{0,2}$ (a), $\delta \nu_{0,1}$, and $\delta \nu_{\rm even,odd}$ (c) small separations. The symbols denote simple LEGACY stars (circles), F-like LEGACY stars (triangles), the non-LEGACY stars (diamonds), Procyon Scenario A (square), and Procyon Scenario B (star). The lines are model tracks for a range of masses 0.9--1.6 M$_\odot$ in steps of 0.1 M$_\odot$, inclusive, of near solar metallicity. The symbol and line colours represents stellar mass shown in (a).}
 \label{fig:figure13}
\end{figure*}
We constructed the standard small separation diagrams \citep[often referred to as C-D diagrams;][]{cd04}, shown in Fig.~\ref{fig:figure13}. We introduce a new variant that makes use of the small `even-odd' separation, $\delta \nu_{\rm even,odd}$ \citep[see][]{bedding10}. The equations of the three small separations are:

\begin{align}
    \label{equ:small_separations}
    \delta \nu_{0,2} &= \nu_{n,0} - \nu_{n-1,2}, \\
    \delta \nu_{0,1} &= \frac{1}{2}(\nu_{n,0} + \nu_{n+1,0}) - \nu_{n,1}, \\
    \delta \nu_{\rm even,odd} &= \frac{1}{2}(\nu_{n,{\rm even}} + \nu_{n+1,{\rm even}}) - \nu_{n,{\rm odd}}.
\end{align}
The small separations can be calculated from the frequency splittings of the ridges shown in Fig.~\ref{fig:figure1}. For $\delta \nu_{0,2}$ and $\delta \nu_{0,1}$ we used the individual mode frequencies from the literature, and for $\delta \nu_{\rm even,odd}$ we used the centroid frequencies from this work.

Considering the model tracks in Fig.~\ref{fig:figure13}, regions where the tracks are spread widely show where that small separation is a good indicator of age. Note that stars evolve from right to left along the tracks. The details of the tracks depend on the chosen value of $\eta$, but do not vary significantly over a range of $\eta$ values. The $\delta \nu_{0,2}$ small separation is known to be a good indicator of age for stars with a wide range of masses. However, it becomes a worse indicator for more evolved stars. The $\delta \nu_{0,1}$ small separation is a fair indicator for some values of $\Delta \nu$ but not others. The $\delta \nu_{\rm even,odd}$ small separation gives a poor indication of age early in the main-sequence, but for more evolved main-sequence stars it becomes a better indicator of age.

We compared the $\delta \nu_{\rm even,odd}$ separation calculated using the ridge centroid peak-bagging against the theoretical small ridge separation. The small ridge separation can be estimated using the conventional small separations and the relative contribution of the respective mode frequencies. Following \cite{bedding10} we have:
\begin{equation}
\label{equ:deo_est}
\delta \nu_{\rm even,odd} = \delta \nu_{0,1} - (1-\eta)\delta \nu_{0,2} + \eta_3 \delta \nu_{1,3},
\end{equation}
where $\eta_3$ is the contribution from the $l{=}3$ mode in the odd ridge centroid. $\eta_3$ can be estimated using the mode visibility of the $l{=}3$ mode. \cite{kjeldsen08b} calculated a value for $\eta_3$ of 0.01 for the \kepler bandpass and 0.11 for radial velocity measurements of Procyon. We neglected the contribution of the $l{=}3$ mode for the stars observed using photometry. For simplicity, we also assumed no contribution of the $l{=}3$ mode for Procyon, mainly because we did not calculate modes of angular degree higher than $l{=}2$ for the grid of models. However, the frequency shift due to the $l{=}3$ mode for the radial velocity data is approximately $0.5$--$1 \mu$Hz, depending on the magnitude of the $\delta \nu_{1,3}$ small separation. $l{=}3$ modes were not explicitly published for Procyon Scenario A, therefore, $\delta \nu_{1,3}$ was not calculated for either ridge identifications. 

\begin{figure}
 \includegraphics[width=1.0\columnwidth]{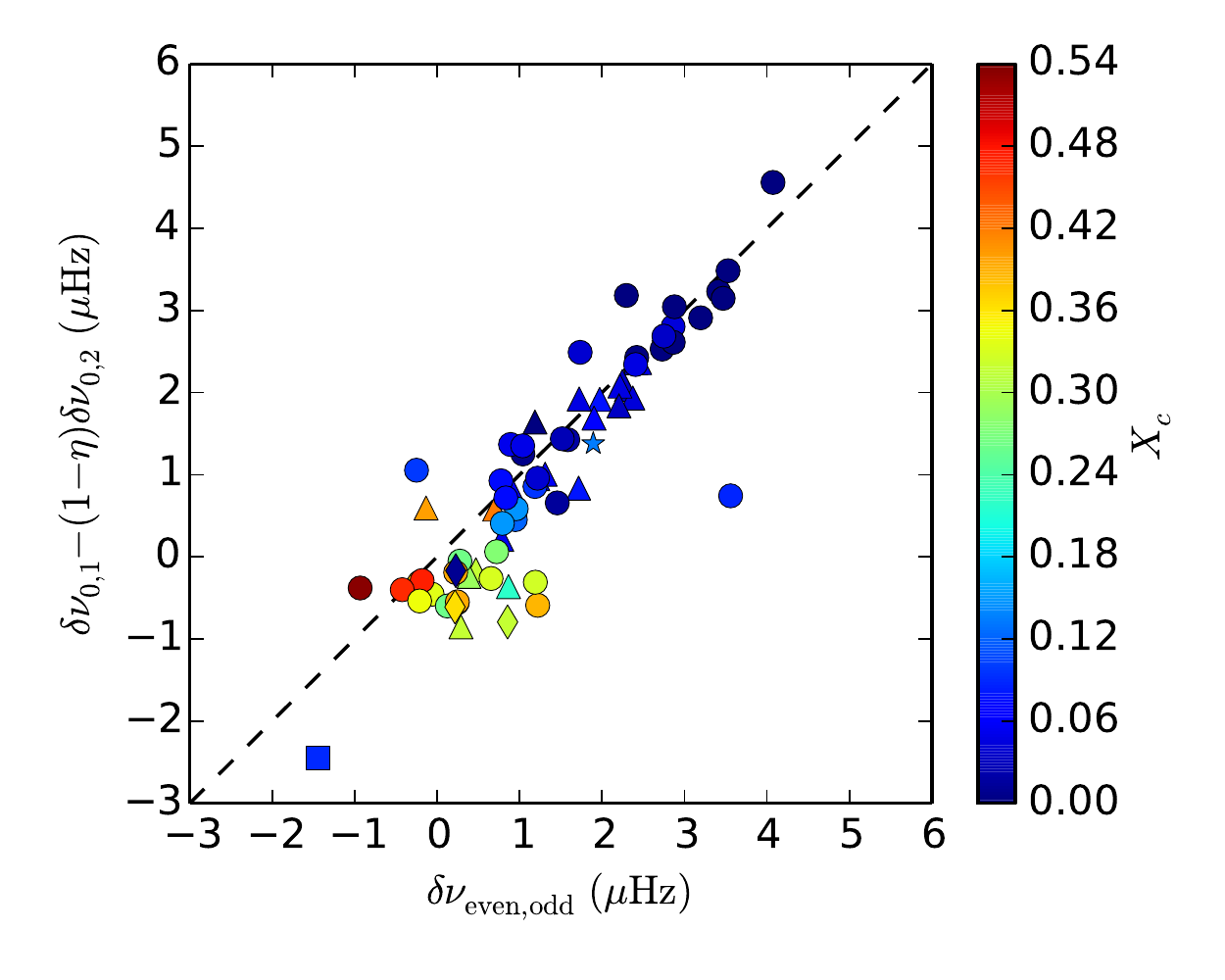}
 \caption{Comparison of $\delta \nu_{\rm even,odd}$ small separation to the other two considered small separations using Eq.~\ref{equ:deo_est}. The symbol shapes represent the same as in Fig.~\ref{fig:figure99} and \ref{fig:figure13}. Colour indicates the measured hydrogen core abundance $X_c$ from the best-fitting models using the \protect\cite{compton18} method with the individual mode frequencies. The dashed line represents equality.}
 \label{fig:figure13b}
\end{figure}
The comparison of the centroid frequency to the small separation frequencies using Eq.~\ref{equ:deo_est} is shown in Fig.~\ref{fig:figure13b}. This suggests that the small ridge separation is a good indicator of stellar age for more evolved main-sequence stars. Note that the clear outlier in the bottom right of Fig.~\ref{fig:figure13b} is the simple star KIC~8424992. Only seven radial orders of modes were used for this star, which is significantly less than the average number.

\subsection{\texorpdfstring{$\theta$}{theta}~Cyg}
\label{sec:theta_cyg}

The asteroseismic analysis of $\theta$~Cyg by \cite{guzik16} used two short-cadence \kepler Quarters. We applied the \cite{compton18} approach using the mode frequencies published by \cite{guzik16}, which returned similar stellar parameters. The ridge centroid analysis returns a mass ($1.37 \pm 0.02$ M$_\odot$) and radius ($1.50 \pm 0.01$ R$_\odot$) that agree with the independent analysis by \cite{white13} ($1.37 \pm 0.04$ M$_\odot$ and $1.48 \pm 0.02$ R$_\odot$). 

\cite{compton18} showed that the surface correction can be used as a diagnostic to determine whether the model is a good fit to the data. However, the value of the relative surface correction at $\numax$ for $\theta$~Cyg disagrees with stars of similar surface gravity calculated by \cite{compton18} when using the individual mode frequencies ($(0.347 \pm 0.832) \cdot 10^{-3}$). Figure~7 in that paper suggests that there is expected to be a non-zero surface correction for $\theta$~Cyg ($\log{g} \simeq 4.23$). The relative surface correction at $\numax$ slightly improves when using the ridge centroid frequencies ($\nu_{\rm corr}(\numax)/\numax = (-1.959 \pm 0.745) \cdot 10^{-3}$) over the underlying ones. The reason for the discrepancy in the surface correction is likely due to inaccurately measured frequencies in $\theta$~Cyg by \cite{guzik16}. Our even centroid frequencies agree remarkably well with a number of the radial-mode frequencies measured by \cite{guzik16}. This indicates that a significant number of the modes measured by \cite{guzik16} were the unresolved centroid frequencies rather than only radial modes.



\subsection{Procyon}
\label{sec:procyon}



For Procyon, Fig.~\ref{fig:figure99} does not show significant differences in modelled stellar parameters between the ridge centroid and individual frequencies with the \cite{compton18} approach. However, the stellar parameters of Procyon disagree with the values from the literature \citep[e.g.][]{girard00,liebert13,guenther14,bond15}. For example, the calculated stellar mass of Procyon Scenario A using the ridge centroid method was estimated at {$1.53 \pm 0.01$ M$_\odot$}, whereas the \emph{Hubble Space Telescope} astrometry of Procyon has shown the mass of $1.478 \pm 0.012$ M$_\odot$ \citep[][]{bond15}. For Procyon Scenario B, the values of stellar mass were similar to the literature, but the radius ($1.98 \pm 0.01$ R$_\odot$) disagrees with the radius measured by \cite{aufdenberg05} using high-precision interferometry ($2.031 \pm 0.013$ R$_\odot$). For both scenarios, the surface correction at $\numax$ does not agree with the values calculated with similar F-type stars.

One explanation is that Procyon is close to the terminal-age main-sequence, as evident by its position in Fig.~\ref{fig:figure2} ($L{\sim}7$~${\rm L}_\odot$, $T_{\rm eff}{\sim}6500$~K), as well as the negative $\delta \nu_{\rm even, odd}$ small separation in Fig~\ref{fig:figure13}. Stars at this stage of evolution can possess mixed modes, which are coupled p and g modes and exhibit characteristics of both \citep[see][]{osaki75,aizenman77,dziembowski91,cd04,aerts10}. Mixed modes strongly deviate from the asymptotic relation and their frequencies evolve rapidly compared to pure p modes. Therefore, fitting mixed modes to a grid of models is inherently difficult, hence the possible mixed mode at 446 $\mu$Hz \citep{bedding10} was neglected in the analysis. The stars in the LEGACY sample were specifically chosen to not contain mixed modes, hence, stars with mixed modes should be omitted for samples like this.



We found that the best way to fit the models to Procyon was to only consider the radial and dipole mode frequencies measured by \cite{bedding10}. This returned a stellar mass of $1.51 \pm 0.02$ M$_\odot$, and a relative surface correction of $(-4.0 \pm 2.3) \cdot 10^{-3}$. Additionally, the radius $R = 2.05 \pm 0.01$ R$_\odot$ agrees with the radius measured by \cite{aufdenberg05} using high-precision interferometry ($R = 2.031 \pm 0.013$ R$_\odot$). Procyon Scenario B returns an underestimated mass ($1.45 \pm 0.01$ M$_\odot$) and radius ($1.94 \pm 0.01$ R$_\odot$), as well as an implausibly large surface correction at $\numax$ of $(-34.7 \pm 2.7) \cdot 10^{-3}$. Therefore, we suggest that Scenario A is the better fitting mode identification based on the radial and dipole modes using the \cite{compton18} method. This conclusion agrees with the asteroseismic analysis of Procyon by \cite{guenther14}, who modelled diffusion and convective overshoot, but disagrees with \cite{white12}, who showed that Scenario B fit better to the $\epsilon$--$T_{\rm eff}$ relation. Upcoming TESS observations \citep{ricker15} will hopefully give a definitive answer to the mode identification problem.


\subsection{HD~49933 and HD~181420}

Finally, we discuss the results of the centroid frequency method on the two CoRoT stars. The model fit to centroid frequencies of HD~49933 returned similar results to the individual frequencies measured by \cite{benomar09} under the \cite{compton18} approach. For both methods, the radius agrees with an interferometric analysis \citep[$1.42 \pm 0.04$ R$_\odot$, see][]{bigot11}. Using the centroid frequencies, the mass of HD~49933 ($1.13 \pm 0.07$ M$_\odot$) agrees with the independent analysis by \cite{bruntt09}, but disagrees with the asteroseismic analyses by \cite{kallinger10} and \cite{liu14}, both estimated a mass of ${\sim}1.3$ M$_{\odot}$ but neglected quadrupole modes in their analysis. Note that \cite{gruberbauer09} and \cite{kallinger10} found no evidence of $l{=}2$ modes in the power spectrum of HD~49933. Therefore, the centroid frequency method appears to be a more robust method of calculating fundamental stellar parameters when the $l{=}0$ and 2 modes cannot be resolved.

For HD~181420, the underlying frequencies \citep[see][]{barban09,gaulme09} performed better than the ridge centroid frequencies with the \cite{compton18} approach. The ridge centroid method produces a slightly underestimated stellar mass with a large uncertainty ($1.27 \pm 0.07$ M$_\odot$) compared to other independent analyses \citep[1.3 -- 1.45 M$_\odot$, see][]{hekker14}. Additionally, the magnitude of the relative surface correction at $\numax$ is more reasonable from the individual mode frequencies ($(-4.2 \pm 3.3) \cdot 10^{-3}$) than the ridge centroid frequencies ($(3.8 \pm 3.2) \cdot 10^{-3}$), primarily because the former is negative.
\section{Conclusions}
\label{sec:conc}

We calculated the odd- and even-degree centroid frequencies for 66 LEGACY stars, $\theta$~Cyg, HD~49933, HD~181420, and Procyon. We implemented a MCMC routine to measure the centroid frequencies of `F-like' and artificially degraded `simple' stars. The centroid frequencies were fitted to a grid of stellar models using a modified \cite{compton18} approach. The surface correction and stellar parameters, calculated using the centroid frequencies, agreed with the individual frequencies for all F-like stars and most simple stars, from the analysis by \cite{compton18} and other published literature. However, the centroid frequency method fails when the observed or modelled data contains mixed modes, therefore care must be taken for stars near the terminal-age main-sequence.

Procyon Scenario A was affected by quadrupole mixed modes and were able to find a model that fit well to Procyon Scenario A by only considering $l{=}0$ and 1 modes published by \cite{bedding10}. The resulting stellar parameters agreed with the literature. In contrast, the fundamental stellar parameters returned for Procyon Scenario B strongly disagreed with the literature. Furthermore, the calculated surface correction for Procyon Scenario B was an extreme outlier compared to every other star in our sample. Therefore, we suggest that Procyon Scenario A is the correct mode identification. For the other non-LEGACY F stars in our sample, we saw generally positive results when using the centroid frequency method.

We introduced the $\delta \nu_{\rm even, odd}$ small separation using the odd- and even-degree ridge centroid frequencies, and found that it is a good indicator for age for evolved main-sequence stars. When fitting modelled oscillation frequencies to observed data, we encourage using the centroid frequencies when the $l{=}0,2$ modes cannot be resolved. This allows some of the power of asteroseismology to be applied to F stars, despite their short mode lifetimes.

\section*{Acknowledgements}
\label{sec:ack}

We thank the referee for helpful comments. This research was supported by the Australian Research Council. Funding for this project has been provided by the Group of Eight Australia-Germany Joint Research Cooperation Scheme. Funding for the Stellar Astrophysics Centre is provided by the Danish National Research Foundation (grant agreement no.: DNRF106). The research is supported by the ASTERISK project (ASTERoseismic Investigations with SONG and Kepler) funded by the European Research Council (grant agreement no.: 267864).

\bibliographystyle{mnras}
\bibliography{ref}

\appendix

\label{lastpage}
\end{document}